\documentclass{article}
\usepackage{spconf,amsmath,graphicx}
\usepackage{bm}
\usepackage{booktabs}
\usepackage{multirow}
\usepackage{tabularx}
\usepackage{comment}
\usepackage{multirow}
\usepackage{url}
\usepackage{amssymb}
\usepackage{pifont}
\usepackage{color}

\newcommand{\gray}[1]{\textcolor[gray]{0.5}{#1}}
\newcommand{\modified}[1]{\textcolor{black}{#1}}

\title{Analysis of Noisy-target Training \\ for DNN-based speech enhancement
\vspace{-5pt}}
\name{Takuya Fujimura, Tomoki Toda
\vspace{-5pt}}
\address{Nagoya University, Japan
\vspace{-8pt}}
\begin{document}
%
\maketitle
\begin{abstract}
Deep neural network (DNN)-based speech enhancement usually uses a clean speech as a training target.
However, it is hard to collect large amounts of clean speech because the recording is very costly.
In other words, the performance of current speech enhancement has been limited by the amount of training data. 
To relax this limitation, Noisy-target Training~(NyTT) that utilizes noisy speech as a training target has been proposed.
Although it has been experimentally shown that NyTT can train a DNN without clean speech, a detailed analysis has not been conducted and its behavior has not been understood well.
In this paper, we conduct various analyses to deepen our understanding of NyTT.
In addition, based on the property of NyTT, we propose a refined method that is comparable to the method using clean speech.
Furthermore, we show that we can improve the performance by using a huge amount of noisy speech with clean speech.
\end{abstract}
\begin{keywords}
Single-channel speech enhancement, deep neural network, Noise2Noise, unsupervised learning
\end{keywords}

\vspace{-2pt}
\section{Introduction}
\vspace{-2pt}
\label{sec:intro}

\begin{figure}[t]
  \centering
\includegraphics[width=0.99\columnwidth,clip]{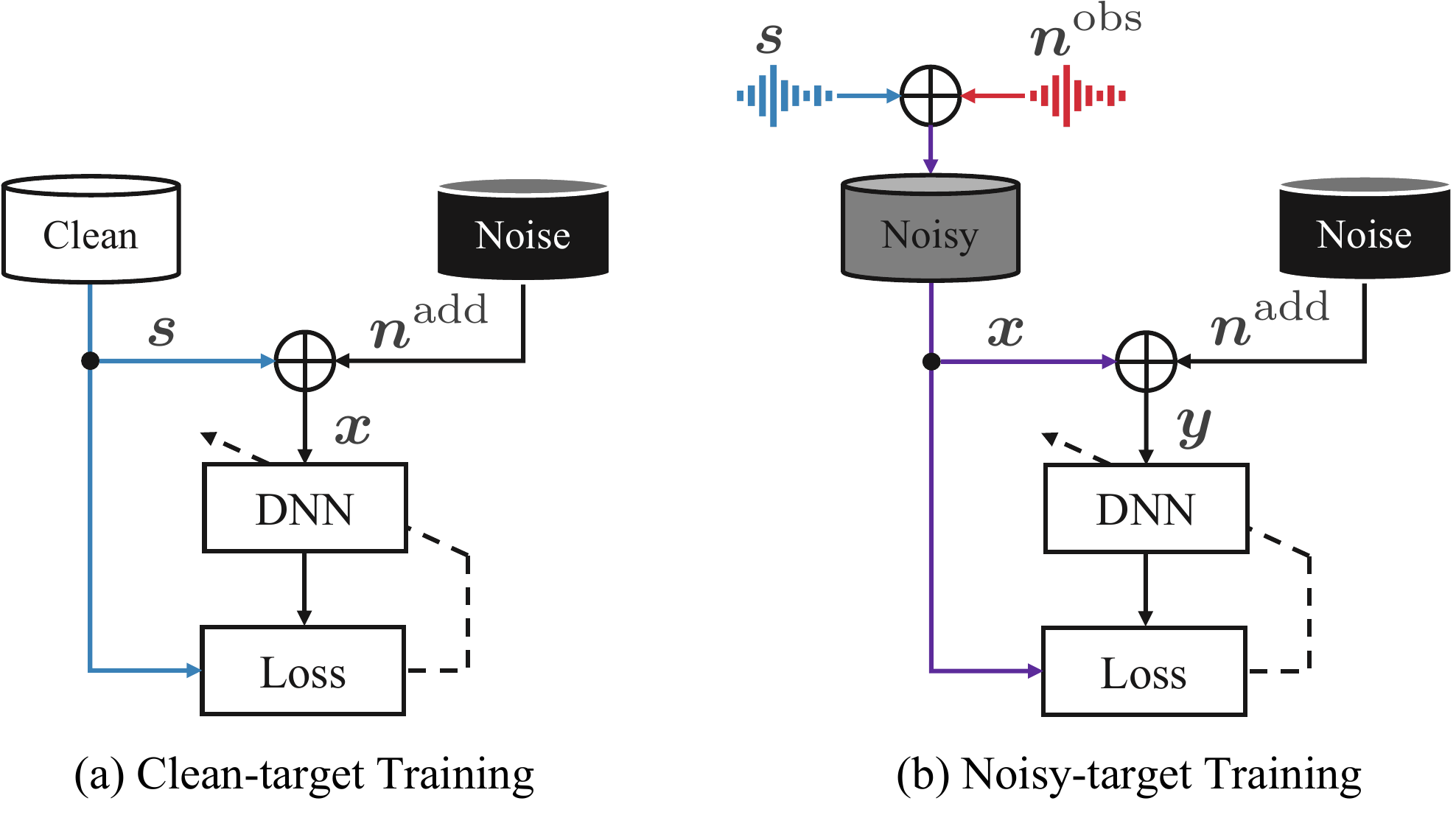} 
\vspace{-15pt}
  \caption{
  Overview of (a) Clean-target Training and (b) Noisy-target Training.
  They are same procedure but Noisy-target Training does not require clean speech signals at all.
  }
  \vspace{-15pt}
\label{fig:methods}
\end{figure}

Speech enhancement aims to extract target speech from noisy observation.
It plays an important role in a variety of applications such as automatic speech recognition (ASR) systems~\cite{Narayanan_2013, Yoshioka_2015, Kinoshita_2020}.
In recent years, speech enhancement has made great advancements by using a deep neural network~(DNN)~\cite{Erdogan_2015, Williamson_2016, luo_2019,Wang_2021,Koizumi_2021}.
Most of the DNN-based speech enhancement method is based on Clean-target Training~(CTT) as shown in Fig.~\ref{fig:methods}~(a).
In CTT, we input noisy speech into a DNN and train it to predict the corresponding clean target speech.
With this approach, many methods have achieved superior performance to \modified{the traditional signal processing methods.} 

Although the CTT is a proper strategy, it has one major problem: \modified{collecting a huge amount of data is hard.}
Because audio signals are easily degraded by environmental noise and reverberation, clean speech signals \modified{should} be recorded in a particular setting such as an anechoic chamber.
It makes the recording very costly and time-consuming and prohibits us from collecting a large-sized dataset. 
This problem has limited the performance of speech enhancement since current deep learning methods perform better with more data~\cite{Brown_2020, Dosovitskiy_2020, Radford_2021}.
Therefore, it is essential to relax the limitation on the amount of data by utilizing signals obtained easily. 

To overcome the limitation, some studies consider approaches that do not require clean speech~\cite{Wisdom_2020, Fu_2022, Fujimura_2021}.
MixIT separates the mixture of two mixtures into each source and trains a DNN to minimize the error between the original mixture and the mixture generated by remixing separated mixtures~\cite{Wisdom_2020}.
MetricGAN-U trains a DNN to improve the speech quality metric, where the metric does not require clean speech~\cite{Fu_2022}.
\modified{A simpler approach than these} is Noisy-target Training~(NyTT) as shown in Fig.~\ref{fig:methods}~(b).
In the NyTT, a DNN predicts a noisy target speech from a \textit{more noisy signal} synthesized by mixing the noisy target speech and an additional noise signal.
In the inference, we input an unprocessed noisy speech and obtain the enhanced speech signal.
In other words, NyTT trains a DNN using noisy speech instead of clean speech.
Here, the model and training procedure are the same for CTT and NyTT.
Therefore, NyTT is important in terms that it can directly benefit from the advanced CTT model, e.g., its performance and processing speed.
Despite its importance, detailed analysis other than experimental proof that this strategy can train DNNs is insufficient.

In this paper, we conduct the analysis of NyTT and deepen our understanding of it.
Our contributions are as follows: (i) proposing a new method based on the property of NyTT and showing its effectiveness,
(ii) clarifying the conditions under which NyTT and its newer version work effectively focusing on the mismatch of noise environment between given data, and
(iii) showing that we can improve the performance of the DNN-based speech enhancement by adding a large-sized dataset of noisy speech to a clean dataset to increase the amount of the training data.

\section{Noisy-target Training}
\label{sec:related}
Noisy-target Training is the training strategy for DNN-based speech enhancement that does not require clean speech signals at all.
NyTT utilizes the observation $\bm{x} = \bm{s} + \bm{n}^{\mathrm{obs}} $ where $\bm{s}$ and $\bm{n}^{\mathrm{obs}}$ denote clean speech and observed noise, respectively.
In the training, we synthesize a \textit{more noisy signal} $\bm{y}$ by mixing a noisy target $\bm{x}$ and an additional noise $\bm{n}^{\mathrm{add}}$ as $\bm{y} = \bm{x} + \bm{n}^{\mathrm{add}}$.
Then, we train a DNN $f(\cdot)$ to minimize the following prediction error $\mathcal{L}$,
\begin{align}
\label{eq:loss}
    \mathcal{L} = \mathbb{E}\left[\mathcal{D}( f(\bm{y}; \theta), \bm{x})\right],
\end{align}
where $\mathbb{E}$ is the expectation operator, $\theta$ is the set of parameters of the DNN $f(\cdot)$, and $\mathcal{D}$ is the mean-squared-error~(MSE). 
A previous study has experimentally shown that NyTT can train a DNN for speech enhancement~\cite{Fujimura_2021}.

\modified{The reason why NyTT works has been believed to be because NyTT can be interpreted as Noise2Noise which is the training strategy for image denoising that does not require clean signals~\cite{Lehtinen_2018}.}
In the Noise2Noise, we train a DNN to learn the mapping from noisy signal $\bm{s}+\bm{n}^{(1)}$ to another noisy signal $\bm{s}+\bm{n}^{(2)}$.
Since a DNN cannot exactly predict random noise components, their expected values are predicted as the noise components when we use MSE as a loss function.
Therefore, by assuming that the noise of the target signal has zero-mean distribution, this training yields a DNN that predicts a clean target signal.
For NyTT, it can be viewed as Noise2Noise with $\bm{y} = \bm{s} + \left(\bm{n}^{\mathrm{obs}}+\bm{n}^{\mathrm{add}}\right)$ and $\bm{s} + \bm{n}^{\mathrm{obs}}$ where the noise signal $\bm{n}^{\mathrm{obs}}$ of the target signal $\bm{x}$ has zero-mean distribution in the time domain.
Thus, we can interpret that the DNN is trained to predict the clean speech signal $\bm{s}$, and we can say that NyTT performs speech enhancement in a Noise2Noise manner.

\section{Motivation and contents of investigation}
\label{sec:motivation}
Although we experimentally showed that NyTT could train a DNN without clean signals~\cite{Fujimura_2021}, its detailed behavior is still unclear.
Therefore, we conduct various investigations of NyTT.
In this section, we summarise the investigation items.

\noindent
\textbf{Validity of interpretation of NyTT:}
So far, NyTT has been interpreted as a Noise2Noise framework.
On the other hand, NyTT can also be interpreted as the removal of $\bm{n}^{\mathrm{add}}$.
Therefore, in Section~\ref{sec:interpretation}, we verify whether NyTT is Noise2Noise or not.
If NyTT was not Noise2Noise, the assumption of zero-mean noise and the use of MSE would be no longer necessary and it enables us to use a variety of loss functions.
Also, it is expected that NyTT will be clearly affected by the mismatch between $\bm{n}^{\mathrm{add}}$ and noise of the test set $\bm{n}^{\mathrm{test}}$.

\noindent
\textbf{Improvement of NyTT by iteration:}
A previous experiment has shown that the performance of NyTT is increased as the SNR of the noisy target increases~\cite{Fujimura_2021}.
Based on this property, we propose an Iterative NyTT (IterNyTT) that achieves higher-quality speech enhancement using a teacher-student learning approach~\cite{Tzinis_2022, Lukas_2019, Xie_2020}.

First, IterNyTT trains a DNN by NyTT with a noisy target dataset.
Next, we apply speech enhancement to the noisy target dataset using the DNN trained in the first step.
Then, IterNyTT trains a new DNN using the enhanced noisy target (which can be viewed as a pseudo-clean target).
Repeating this process is expected to improve the SNR of noisy targets, and thus the speech enhancement performance.
We verifies the effectiveness of IterNyTT in Section~\ref{sec:iterNyTT}.

\noindent
\textbf{Effects of mismatch between each noise:}
In the NyTT, there are three noise data: $\bm{n}^{\mathrm{obs}}$, $\bm{n}^{\mathrm{add}}$, and $\bm{n}^{\mathrm{test}}$.
If NyTT was the removal of $\bm{n}^{\mathrm{add}}$, it could not work when there was a mismatch between $\bm{n}^{\mathrm{add}}$ and $\bm{n}^{\mathrm{test}}$.
In Section~\ref{sec:noise}, we investigate the various effects of mismatch among these noise on NyTT.

\noindent
\textbf{Evaluation using a large-sized noisy dataset:}
It is difficult to improve speech enhancement performance by increasing clean data due to its recording costs.
In contrast, we can obtain noisy signals anywhere.
In Section~\ref{sec:scaleup}, we practically investigate whether we can improve the performance by using a large-sized dataset of noisy speech.

\vspace{-5pt}
\section{Experiments}
\vspace{-5pt}
\label{sec:exp}

\subsection{Experimental setups}
\label{sec:setup}
\vspace{-2pt}
In the experiments, we used LibriSpeech~\cite{libri} as clean speech and CHiME3~\cite{chime3}, the training noise dataset of VoiceBank-DEMAND~\cite{vbd}, and the train set of DCASE 2016 Challenge Task2 (\texttt{DCASE})~\cite{dcase} as noise signals.
The 10,000 and 90,000 utterances from train set of LibriSpeech were used as \texttt{Libri1} and \texttt{Libri2}, respectively.
The background noise data of CHiME3 was cut every 10 seconds and 7.83 hours of that was divided into two sets: \texttt{CHiME1} and \texttt{CHiME2}.
Also, the training noise dataset of VoiceBank-DEMAND was divided into two sets: \texttt{DEMAND1} and \texttt{DEMAND2}.
The test set was generated by mixing 1,000 utterances from the test set of LibriSpeech and 0.56 hours of CHiME3 (\texttt{CHiME}) at SNR randomly selected from 2.5, 7.5, 12.5, and 17.5~dB.
The noisy target dataset was generated by mixing clean speeches and noise signals $\bm{n}^{\mathrm{obs}}$ at SNR randomly selected from 0, 5, 10, and 15~dB.
We refer to this dataset as ``clean set name\texttt{-}noise set name'', i.e., \texttt{Libri1-CHiME1} means a mixture set of \texttt{Libri1} and \texttt{CHiME1}.
In Section \ref{sec:interpretation} and \ref{sec:iterNyTT}, we used \texttt{CHiME1} as $\bm{n}^{\mathrm{add}}$ and \texttt{Libri1-CHiME1} as noisy target $\bm{x}$.

The DNN estimating a complex-valued T-F mask~\cite{Williamson_2016} consisted of CNN-BLSTM used in~\cite{Kawanaka_2020}.
The input feature was log-amplitude spectrogram of the input signal.
For the short-time Fourier transform parameters, frame shift, window size, and DFT size were set to 128-, 512-, and 512-samples with the Hamming window, where the sampling frequency was 16~kHz.
We trained the DNN 200 epochs with mini-batch size 50 using Adam optimizer~\cite{Kingma_2015} with a fixed learning rate 0.0001.
For a loss fucntion, we used MSE calculated in the time domain.
For NyTT, we generated more noisy signals by mixing noisy target $\bm{x}$ and noise $\bm{n}^{\mathrm{add}}$ at SNR randomly selected between -5 to 5~dB, where SNR was measured by considering $\bm{x}$ as the target signal and $\bm{n}^{\mathrm{add}}$ as noise.
For a comparison method CTT and IterNyTT on or after the second iteration, we generated the input signals by mixing clean or enhanced noisy target and noise $\bm{n}^{\mathrm{add}}$ at SNR randomly selected from 0, 5, 10, and 15~dB.

As the metrics, we used scale-invariant signal-to-distortion ratio (SI-SDR) and PESQ.
These are the standard metrics for evaluation of speech enhancement.

\begin{table}[tt]
\caption{\modified{Evaluation of the signals treated in NyTT}}
\label{tbl:morenoisy}
\begin{center}
\vspace{0pt}
\footnotesize
\begin{tabular}{c|c|cccc} 
\toprule
   Data & Metric & Noisy target & More noisy & Output\\ \midrule
\multirow{2}{*}{Train set} & SI-SDR & \textbf{8.33} & 1.37 & 6.73\\
    & PESQ & \textbf{1.34} & 1.07 & 1.28\\ \midrule
\multirow{2}{*}{Test set} & SI-SDR & \textbf{10.69} & 1.72 & 7.49\\
    & PESQ & \textbf{1.53} & 1.08 & 1.33\\
\bottomrule
  \end{tabular}
\end{center}
\vspace{-20pt}
\end{table}

\begin{table}[tt]
\caption{Comparison of a loss function.}
\label{tbl:interpretation}
\begin{center}
\footnotesize
\begin{tabular}{l|cccc} 
\toprule
     & Input & \texttt{Time}  & \texttt{Spec}\\ \midrule
    SI-SDR & 10.69 & \textbf{15.03} & 14.78\\
    PESQ & 1.53 & \textbf{2.23} & 2.09\\
\bottomrule
  \end{tabular}
\end{center}
\vspace{-22pt}
\end{table}

\begin{figure}[t]
\centering
\centerline{\includegraphics[width=\linewidth]{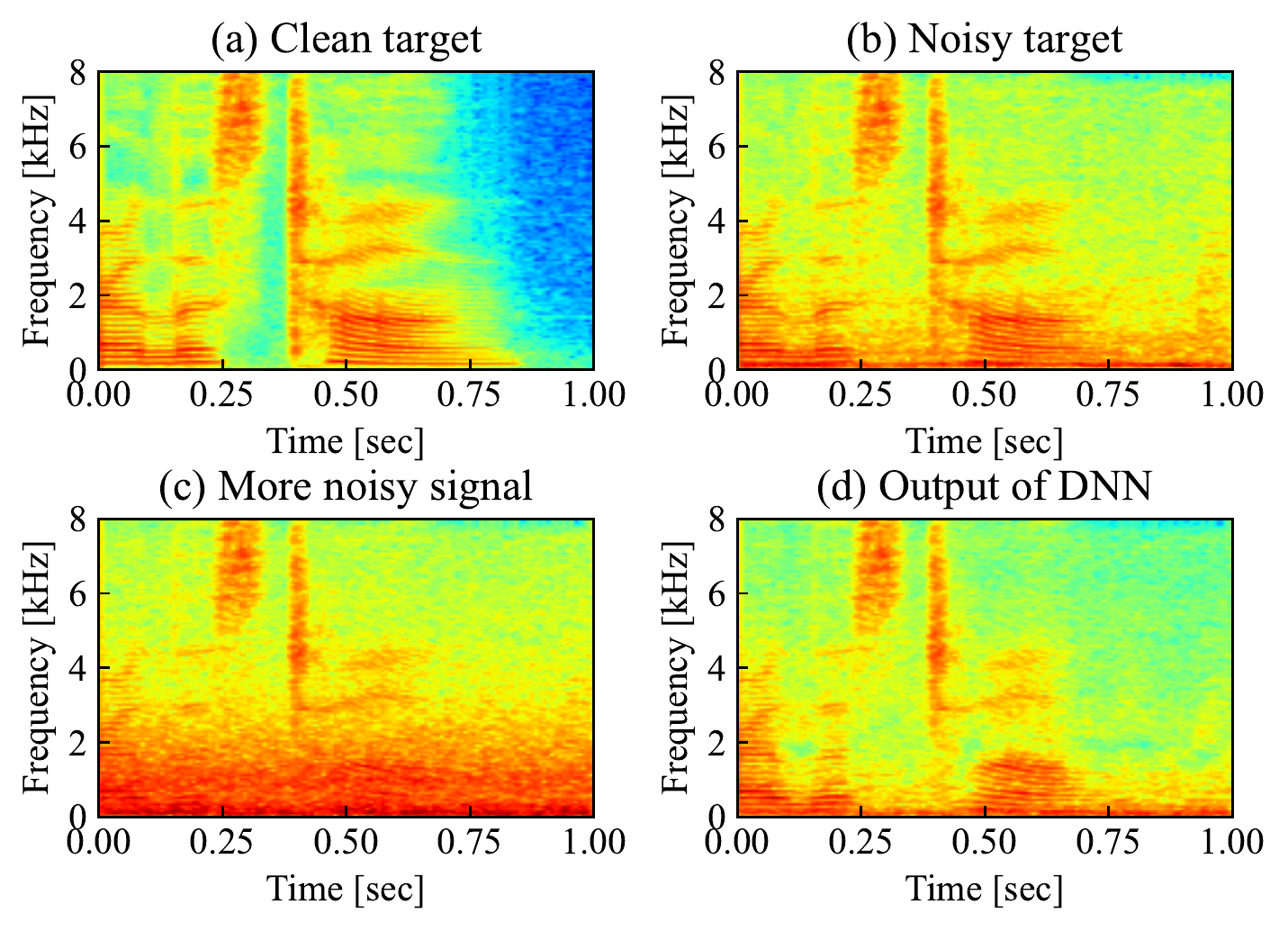}}
\vspace{-10pt}
\caption{Spectrograms of testing data. (a) Clean target, (b) noisy target, (c) more noisy signal, and (d) output of a DNN.}
\label{fig:spec}
\vspace{-9pt}
\end{figure}

\vspace{-5pt}
\subsection{Validity of interpretation of NyTT}
\vspace{-3pt}
\label{sec:interpretation}
In this section, we conducted two analyses to confirm whether NyTT is Noise2Noise.
First, we analyzed the output signal of the DNN when we input the more noisy signal.
If NyTT was Noise2Noise, the output signal would be the estimate of clean speech $\bm{s}$.
In contrast, if NyTT was the removal of $\bm{n}^{\mathrm{add}}$, the output signal would be the estimate of noisy target $\bm{x}$.
\modified{For this analysis, we generated the more noisy signals using training and testing data, respectively.
More noisy signals of the train set were generated using 1,000 utterances from \texttt{Libri1-CHiME1} as the noisy target and \texttt{CHiME2} as $\bm{n}^{\mathrm{add}}$, where they were used during the training.
More noisy signals of the test set were generated using 1,000 utterances of the testing noisy data and \texttt{CHiME}.}
Table~\ref{tbl:morenoisy} shows the evaluated scores of the noisy targets $\bm{x}$, more noisy signals $\bm{y}$, and output signals $f(\bm{y};\theta)$ \modified{of the both train and test set.}
The output is closer to the noisy target than the clean target, considering its speech quality.
An example of a spectrogram is shown in Fig.~\ref{fig:spec}.
The figure also suggests that the output of a DNN was an estimate of a noisy target rather than that of a clean target.

Second, we compared the performance when the loss function, MSE, was calculated in the time domain (\texttt{Time}) with the performance when it was calculated in the amplitude spectrogram domain (\texttt{Spec}).
If NyTT was not Noise2Noise, \texttt{Spec} could also perform speech enhancement, even though it cannot assume zero-mean noise.
In this experiment, the DNN estimated the real T-F mask to be applied to the amplitude spectrogram, and the phase of the input signal was used to convert it to the time domain.
Table~\ref{tbl:interpretation} shows both \texttt{Time} and \texttt{Spec} conditions improved the evaluated scores of the input signal.
This result indicates that NyTT worked without the assumption about zero-mean noise and MSE.

All of the above results indicate that NyTT is rather interpreted as the training to predict noisy target $\bm{x}$ by removing $\bm{n}^{\mathrm{add}}$ than the training of Noise2Noise.

\vspace{-5pt}
\subsection{Effectiveness of IterNyTT}
\begin{figure}[t]
\centering
\includegraphics[width=0.99\columnwidth,clip]{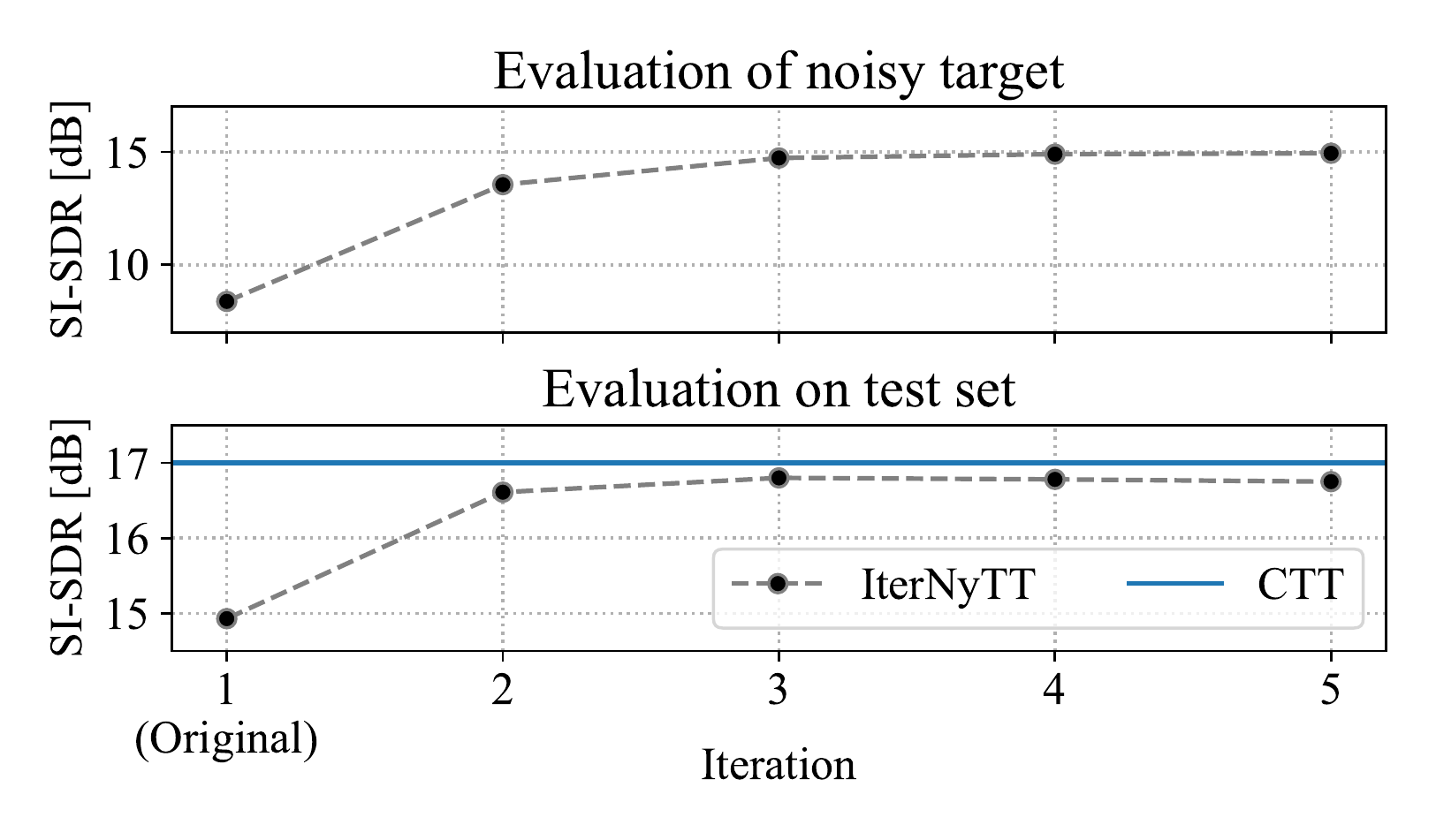}
\vspace{-18pt}
\caption{Relationship between iteration and SI-SDR of noisy target and evaluation result of IterNyTT. The first iteration of IterNyTT is equivalent to the normal NyTT.}
\label{fig:iterNyTT}
\vspace{-12pt}
\end{figure}

\vspace{-3pt}
\label{sec:iterNyTT}
In this experiment, we investigated whether the performance of IterNyTT improves during five iterations.
Figure~\ref{fig:iterNyTT} shows the experimental results.
IterNyTT improved SI-SDR of the noisy target, and the score of IterNyTT approached that of CTT as the iteration increased.
We can also see that IterNyTT already achieved a high score at the second iteration.
In the following analyses, we set the iteration of IterNyTT to three.

\vspace{-8pt}
\subsection{Effects of mismatch between each noise}
\label{sec:noise}
\vspace{-3pt}
In this section, we investigated the effects of the mismatch of noise on NyTT.
In the experiment, we used \texttt{CHiME1}, \texttt{DEMAND1}, and \texttt{DCASE} as $\bm{n}^{\mathrm{obs}}$ and \texttt{CHiME2} and \texttt{DEMAND2} as $\bm{n}^{\mathrm{add}}$.
The noisy target consisted of \texttt{Libri1} and each $\bm{n}^{\mathrm{obs}}$ data.
Table~\ref{tbl:noise} summarizes the SI-SDR in each combination for $\bm{n}^{\mathrm{obs}}$ and $\bm{n}^{\mathrm{add}}$ where $\bm{n}^{\mathrm{test}}$ is \texttt{CHiME}.
We analyzed this result from three perspectives: mismatches between $\bm{n}^{\mathrm{add}}$ and $\bm{n}^{\mathrm{test}}$, $\bm{n}^{\mathrm{add}}$ and $\bm{n}^{\mathrm{obs}}$, and 
$\bm{n}^{\mathrm{obs}}$ and $\bm{n}^{\mathrm{test}}$.
First, for each $\bm{n}^{\mathrm{obs}}$, we can see that higher scores were obtained when $\bm{n}^{\mathrm{add}}$ was \texttt{CHiME2} which matches $\bm{n}^{\mathrm{test}}$.
Second, the table shows that IterNyTT improved the SI-SDR by more than 1.3~dB from NyTT when there was no mismatch between $\bm{n}^{\mathrm{add}}$ and $\bm{n}^{\mathrm{obs}}$.
On the other hand, there was no or slight improvement when there was a mismatch.
\modified{To further analyze the effectiveness of IterNyTT in consideration of practical use, we evaluated the IterNyTT that switches $\bm{n}^{\mathrm{add}}$ according to the removal of the $\bm{n}^{\mathrm{obs}}$ and the $\bm{n}^{\mathrm{test}}$.}
In the case shown at the bottom of the table, we started IterNyTT using \texttt{DEMAND2} for $\bm{n}^{\mathrm{add}}$ and finished it using \texttt{CHiME2} for $\bm{n}^{\mathrm{add}}$ in the third iteration, \modified{where \texttt{DEMAND2} matches \texttt{DEMAND1} of $\bm{n}^{\mathrm{obs}}$ and \texttt{CHiME2} matches \texttt{CHiME} of $\bm{n}^{\mathrm{test}}$.}
In this way, we could obtain comparable performance to CTT and IterNyTT of the situation where $\bm{n}^{\mathrm{obs}}$ and $\bm{n}^{\mathrm{add}}$ were \texttt{CHiME1} and \texttt{CHiME2}.
Finally, we focus on the mismatch between $\bm{n}^{\mathrm{obs}}$ and $\bm{n}^{\mathrm{test}}$.
For each $\bm{n}^{\mathrm{add}}$, we can see that the lowest scores were obtained when $\bm{n}^{\mathrm{obs}}$ was \texttt{CHiME1}.
We believe this result was caused by the training of the DNN to include $\bm{n}^{\mathrm{obs}}$ which matches $\bm{n}^{\mathrm{test}}$ in the output.

Summing up the above results, in the framework of NyTT, (i) the performance will be degraded similarly to CTT when there is a mismatch between $\bm{n}^{\mathrm{add}}$ and $\bm{n}^{\mathrm{test}}$,
(ii) IterNyTT is comparable to CTT when there is no mismatch between $\bm{n}^{\mathrm{add}}$ and $\bm{n}^{\mathrm{obs}}$,
and (iii) the performance will be degraded when there is no mismatch between $\bm{n}^{\mathrm{obs}}$ and $\bm{n}^{\mathrm{test}}$.
In addition, these results are also consistent with the interpretation that NyTT predicts noisy target $\bm{x}$ by removing $\bm{n}^{\mathrm{add}}$.

\begin{table}[tt]
\vspace{-11pt}
\caption{\modified{Results of SI-SDR in each noise combination.}
Input noisy signals consisted of \texttt{CHiME} and its SI-SDR was 10.69.}
\vspace{-5pt}
\label{tbl:noise}
\begin{center}
\footnotesize
\begin{tabular}{cc|ccccc}
\toprule
$\bm{n}^{\mathrm{obs}}$ & $\bm{n}^{\mathrm{add}}$ & CTT & NyTT & IterNyTT\\
\midrule
\texttt{CHiME1} & \texttt{CHiME2} & \textbf{17.00} & 14.93 & 16.80\\
\texttt{CHiME1} & \texttt{DEMAND2} & \textbf{14.83} & 11.08 & 10.73\\
\texttt{DEMAND1} & \texttt{CHiME2} & \textbf{17.00} & 15.65 & 16.56\\
\texttt{DEMAND1} & \texttt{DEMAND2} & \textbf{14.83} & 13.48 & 14.81\\
\texttt{DCASE} & \texttt{CHiME2} & \textbf{17.00} & 16.19 & 16.76\\
\texttt{DCASE} & \texttt{DEMAND2} & \textbf{14.83} & 14.64 & 14.31\\
\midrule \texttt{DEMAND1} & \begin{tabular}{c}\texttt{CHiME2}, \\ \texttt{DEMAND2}\end{tabular} & -- & -- & 16.88\\
\bottomrule
\vspace{-25pt}
\end{tabular}
\end{center}
\end{table}

\vspace{-7pt}
\subsection{Evaluation using a large-sized noisy dataset}
\vspace{-3pt}
\label{sec:scaleup}
In this section, we conducted two evaluations using a large-sized dataset of noisy speech.
First, we confirmed whether NyTT and IterNyTT using noisy \texttt{Libri2} outperformed CTT using \texttt{Libri1}, where the amount of \texttt{Libri2} is nine times that of \texttt{Libri1}.
Table~\ref{tbl:scaleup1} summarizes the SI-SDR, \modified{where IterNyTT with \texttt{Libri2-DEMAND1} switched $\bm{n}^{\mathrm{add}}$ between \texttt{DEMAND2} and \texttt{CHiME2} according to $\bm{n}^{\mathrm{obs}}$ and $\bm{n}^{\mathrm{test}}$.}
From the table, we can see that even normal NyTT outperformed CTT when there was a mismatch between $\bm{n}^{\mathrm{obs}}$ and $\bm{n}^{\mathrm{test}}$ (\texttt{CHiME}).
IterNyTT improved the performance and approached the score of the ideal case (CTT using \texttt{Libri2}).

Second, we discuss a situation where not only a large-sized noisy dataset but also a clean dataset \texttt{Libri1} are available.
In this case, the performance can be improved by using both noisy and clean speech as targets.
For the joint use of clean and noisy targets, we have two options.
The first is simple: use original noisy targets with clean targets.
The second is to use enhanced noisy data.
If we had $\bm{n}^{\mathrm{add}}$ corresponding to $\bm{n}^{\mathrm{obs}}$ of noisy data, we could enhance it by CTT that used \texttt{Libri1} and the $\bm{n}^{\mathrm{add}}$.
Therefore, we also considered the option of using enhanced noisy data.
Table~\ref{tbl:scaleup2} summarizes the SI-SDR and PESQ where the joint use improved performance even without speech enhancement of the noisy target.
However, we can see that adding \texttt{Libri2-CHiME1} was less effective because there was no mismatch between $\bm{n}^{\mathrm{obs}}$ and $\bm{n}^{\mathrm{test}}$.
Furthermore, the table shows that adding the enhanced noisy target was very effective.
These results suggest that we can scale up the training data and the performance of DNN-based speech enhancement by utilizing a huge amount of noisy data.

\begin{table}[tt]
\vspace{-11pt}
\caption{\modified{Results of SI-SDR when} using a large-sized noisy dataset. Input signals consisted of \texttt{CHiME} and its SI-SDR was 10.69. 
We assume that we cannot access to \texttt{Libri2}.}
\vspace{-3pt}
\label{tbl:scaleup1}
\begin{center}
\footnotesize
\begin{tabular}{c|cccc}
\toprule
$\bm{n}^{\mathrm{obs}}$ &  \begin{tabular}{c}CTT \\ (\texttt{Libri1})\end{tabular} & NyTT & IterNyTT & \begin{tabular}{c}\gray{CTT} \\ \gray{(\texttt{Libri2})}\end{tabular}\\
\midrule
\texttt{CHiME1} & 17.00 & 16.62 & \textbf{17.51} & \gray{17.96}\\
\texttt{DEMAND1} &  17.00 & 17.27 & \textbf{17.79} & \gray{17.96}\\
\bottomrule
\end{tabular}
\end{center}
\vspace{-20pt}
\end{table}

\begin{table}[tt]
\caption{Evaluation using both clean and noisy targets.
Input signals consisted of \texttt{CHiME} and its SI-SDR and PESQ were 10.69 and 1.53.
We assume that we cannot access to \texttt{Libri2}.}
\vspace{-2pt}
\label{tbl:scaleup2}
\begin{center}
\footnotesize
\begin{tabular}{cc|cc}
\toprule
Speech datasets & enhanced? & SI-SDR & PESQ \\
\midrule
\texttt{Libri1} & -- & 17.00 & 2.46 \\
\texttt{Libri1}, \texttt{Libri2-CHiME1} & No & 17.03 & 2.47\\
\texttt{Libri1}, \texttt{Libri2-CHiME1} & Yes & 17.86 & 2.62\\
\texttt{Libri1}, \texttt{Libri2-DEMAND1} & No & 17.57 & 2.59 \\
\texttt{Libri1}, \texttt{Libri2-DEMAND1} & Yes & \textbf{17.87} & \textbf{2.67} \\
\gray{\texttt{Libri1}, \texttt{Libri2}} & -- & \gray{18.05} & \gray{2.74} \\
\bottomrule
\end{tabular}
\end{center}
\vspace{-20pt}
\end{table}

\vspace{-7pt}
\section{Conclusion}
\vspace{-6pt}
In this study, we conducted various analyses of NyTT to reveal its behavior.
Our experiments showed that (i) NyTT could be interpreted as training to predict the noisy target by removing $\bm{n}^{\mathrm{add}}$, 
(ii) IterNyTT was comparable to CTT when $\bm{n}^{\mathrm{add}}$ corresponding to $\bm{n}^{\mathrm{obs}}$ was available, 
(iii) the mismatch between $\bm{n}^{\mathrm{obs}}$ and $\bm{n}^{\mathrm{test}}$ was preferable, 
and (iv) we could improve the performance of CTT by adding a large-sized noisy dataset.
In addition, the experiment suggests that NyTT can be applied to any CTT method without the terms of zero-mean noise and MSE because NyTT was not Noise2Noise.
\vspace{3pt}

\noindent
\textbf{Acknowledgements:} This work was partly supported by JST CREST Grant Number JPMJCR19A3, and JSPS KAKENHI Grant Number JP20H00102.

\setlength{\itemsep}{-2pt}
\setlength{\baselineskip}{10pt}


\end{document}